\documentclass[useAMS,usenatbib]{mn2e}

\title[Variation of $\alpha$ from dark matter oscillations]{Variation of 
the fine structure constant in QSO spectra from coherent dark matter 
oscillations}
\author[M. G. Barnett, R. Dick \& K. E. Wunderle]{M. G. Barnett,
 R. Dick 
 and K. E. Wunderle\\
Department of Physics and Engineering Physics,
University of Saskatchewan, Saskatoon, Canada SK S7N 5E2}
\begin{document}

\date{Accepted ---. Received ---; in original form ---}

\pagerange{\pageref{firstpage}--\pageref{lastpage}} \pubyear{2004}

\maketitle

\label{firstpage}

\begin{abstract}
We consider the problem of the evolution of the
fine structure coefficient $\alpha$ under the assumption
that the scalar field coupling to the Maxwell
term satisfies the condition $mt\gg 1$ for coherent
dark matter oscillations.\\
In this case we find that the coupling scale $f$
in the leading order coupling $-(\phi/4f)F_{\mu\nu}F^{\mu\nu}$
affects the cosmological evolution of $\alpha$ according to
$\ln(\alpha/\alpha_0)\propto\xi(m_{Pl}/f)\times\ln(\tanh(t/2\tau)/\tanh(t_0/2\tau))$.
A fit to the QSO observations by Murphy et al. yields 
$f=\xi\times 2.12^{+0.58}_{-0.37}\times 10^5m_{Pl}$. 
Here $m_{Pl}=(8\pi G_N)^{-1/2}$ is the reduced Planck mass,
and $\xi^2=\varrho_\phi/\varrho_m$ parametrizes 
the contribution of $\phi$ to the matter density in the universe.
\end{abstract}

\begin{keywords}
atomic data -- cosmology: theory -- dark matter.
\end{keywords}

\section{Introduction}\label{sec:intro}

The question whether the value of Sommerfeld's fine structure
constant should actually be determined through the dynamics
of a scalar field had been addressed already by \citet{fierz}
and \citet{jordan}, whose investigations were
partly motivated by Kaluza--Klein theory
and by Dirac's proposal of a variability of constants
over cosmological time scales.
Nowadays it is well known that dynamical gauge couplings are
predicted by string theory, and the relevant
coupling e.g. of the heterotic string dilaton to gauge fields
in four dimensions is of gravitational strength
$f^{-1}=(16\pi G_N)^{1/2}=\sqrt{2} m_{Pl}^{-1}$ \citep{rdfp}.
In an independent development
\citet{jacob1} had introduced a class of models
for dynamical $\alpha$ where the evolution of the fine structure constant
is driven through couplings to energy densities.

Therefore there was always theoretical interest in dynamical models for 
$\alpha$, but in recent years Webb et al. also reported 
evidence for a variation of the fine structure coefficient
over cosmological time scales \citep{webb1,murphy1,webb2,murphy2}.
The analysis of 128 quasar absorption systems by \citet{murphy2}
found an average increase\footnote{We follow the standard sign convention
$\Delta\alpha\equiv\alpha_z-\alpha_0$, such that $\Delta\alpha<0$
corresponds to an increase of $\alpha$ with time.}
\begin{equation}\label{eq:webb}
\frac{\Delta\alpha}{\alpha}\equiv \frac{\alpha_z}{\alpha}-1
=-(0.543\pm 0.116)\times 10^{-5},
\end{equation}
since redshift $z=1.67$, i.e. over the last
$\approx 9.6$ billion years.
Here $\alpha_z$ is the fine structure coefficient at redshift $z$
and $\alpha\equiv\alpha_0\equiv\alpha(t_0)$. In \citet{murphy3}
results are reported for 143 quasar apsorption lines, yielding
$\Delta\alpha/\alpha=-(0.57\pm 0.11)\times 10^{-5}$ since $z=1.75$,
but in our present analysis we used the well documented sample from
\citet{murphy2}.

Dynamical gauge couplings can equivalently be expressed as dynamical
permeabilities, see e.g. \citet{MSK}.
Suppose $q$ is a particular fixed
value for the dynamical gauge coupling $Q(x)$ ($q$ will be further
specified below).
With the transformation $qA_\mu(x)=Q(x)\mathcal{A}_\mu(x)$
the covariant derivatives can be written in terms of
a variable or a constant gauge coupling
\[
D_\mu(x)
=\partial_\mu-\mathrm{i}Q(x)\mathcal{A}_\mu(x)
=\partial_\mu-\mathrm{i}qA_\mu(x),
\]
but in the theory with the manifestly variable 
gauge coupling $Q(x)$ the field strength tensor is
\[
\mathcal{F}_{\mu\nu}=\frac{\mathrm{i}}{Q}[D_\mu,D_\nu]
=\frac{1}{Q}
\partial_\mu(Q\mathcal{A}_\nu)-\frac{1}{Q}\partial_\nu(Q\mathcal{A}_\mu)
\]
while the field strength tensor with the constant coupling has the
standard form $F_{\mu\nu}=\partial_\mu A_\nu-\partial_\nu A_\mu
=(Q/q)\mathcal{F}_{\mu\nu}$. As a consequence the gauge theory with 
variable coupling
and constant permeability appears as a gauge theory with
constant coupling and variable permeability:
\[
-\frac{1}{4\mu_0}\mathcal{F}_{\mu\nu}(x)\mathcal{F}^{\mu\nu}(x)
=-\frac{q^2}{4\mu_0 Q^2(x)}F_{\mu\nu}(x) F^{\mu\nu}(x).
\]

The dynamical coupling constant for
charge $Ze$ and the dynamical permeability in SI units are
\[
Q(x)=Ze(x)/\hbar=2Z\sqrt{\pi\epsilon_0 c\alpha(x)/\hbar}
\] 
and
\[
\mu(x)=\mu_0\alpha(x)/\alpha_0=\frac{4\pi\hbar}{ce_0^2}\alpha(x),
\]
respectively. In the sequel we use units with $\hbar=c=1$.

The scalar variable $Q(x)$ may not have a canonically normalized
kinetic term. Therefore a transformation $Q(x)=Q(\phi(x))$
may be required if we want in leading order
a standard $(\partial\phi)^2$ term for the
dynamics of $Q$:
\begin{equation}\label{eq:L}
\mathcal{L}=-\frac{q^2}{4Q^2(\phi)}F_{\mu\nu} F^{\mu\nu}
-\frac{1}{2}\partial_\mu\phi\partial^\mu\phi-V(\phi).
\end{equation}

Here we assume that 
the potential $V(\phi)$ has a unique minimum
$V(\phi_0)$ at some value $\phi=\phi_0$, and we parametrize the scalar
field $\phi$ such that $\phi_0=0$. Furthermore, any nonvanishing
term $V(0)$ would contribute to the cosmological constant and will not
be considered as part of the energy density $\varrho_\phi$
stored in the scalar field $\phi$.
The leading order expansion of the potential is then
\begin{equation}\label{eq:V}
V(\phi)\simeq \frac{1}{2}m^2\phi^2,
\quad
m^2=\left.\frac{d^2V(\phi)}{d\phi^2}\right|_{\phi=0}.
\end{equation}
With this proviso it seems prudent to choose $q=Q(0)$
as the equilibrium value to which the gauge coupling should evolve
due to the presence of the Hubble term $3H\dot{\phi}$. 
In leading order this implies the following parametrization for
the coupling function $Q(\phi)$:
\begin{equation}\label{eq:Qlin}
Q^2(\phi)\simeq q^2\left(1-\frac{\phi}{f}\right),
\end{equation}
with the coupling scale defined accordingly
\begin{equation}\label{eq:deff}
\frac{1}{f}\equiv -\frac{1}{q^2}
\left.\frac{dQ^2(\phi)}{d\phi}\right|_{\phi=0}.
\end{equation}

Besides the convention $\phi_0=0$ for the equilibrium position
this also implies a sign convention on the field $\phi$ if we 
require $f>0$:
$\phi$ is chosen as positive if it reduces the fine structure
constant in first order (see also e.g. \citet{DN}), 
and the observations of Murphy et al.
then indicate that $\phi$ is decaying from a positive value towards its
equilibrium value $\phi_0=0$.

Examples of specific coupling functions
are provided e.g. by string theory or Kaluza--Klein theories:
\begin{equation}\label{eq:stringQ}
Q^2(\phi)=q^2\exp(-\phi/f),
\end{equation}
and the Coulomb problem in these theories exhibits 
an ultraviolet regularization at a scale\footnote{Subsequently
the abelian
and non-abelian Coulomb problem was also found to be exactly solvable
for other functions $Q(\phi)$ \citep{rdepj,chabab,SW},
and with mass couplings of $\phi$ \citep{rdplb3,jacob2}.}
$r_f=q/(8\pi f)$ \citep{rdplb1}.

 For the present investigation we will not specify
the coupling function $Q(\phi)$ any further but only
use the linear expansion (\ref{eq:Qlin}).

\citet{LV} reconsidered the original Bekenstein model and 
concluded that it would not comply with the observations of Murphy et al.
In a very interesting extension \citet{OP1} investigated a model where 
the dynamics of a massless scalar field $\phi$ is mostly driven by its 
couplings to dark matter and the cosmological constant, and
they analyzed compatibility of (\ref{eq:webb}) 
with various constraints on variations of $\alpha$. 
Our coupling parameter $f$ is related to the parameters 
$M_\ast$, $\zeta_F$ and $\omega$ in \citet{OP1} through
$f=M_\ast/\zeta_F=\sqrt{2\omega}M_{Pl}/\zeta_F
=4\sqrt{\pi\omega}m_{Pl}/\zeta_F$, 
and their results favor $f>10^3 M_{Pl}$, corresponding to a 
subgravitational coupling strength of $\phi$ to photons.

\citet{gardner} has recently discussed the implications
of a mass term on the evolution of the fine-structure constant,
and reported it to be consistent with mass values
for the scalar field $\phi$ around $m\simeq H_0\simeq 10^{-33}\,$eV. 
Three crucial assumptions in Gardner's work are that the contribution of 
$\phi$ to the dark matter density is negligible, that $f\le m_{Pl}$, and 
that $0<|\zeta_m| m_{Pl}^2/f^2< 10^{-5}$, where $\zeta_m$ is the
coupling of $\phi$ to matter. A low mass value $m\sim H_0$ was also 
preferred in the recent work by \citet{luis}, who identified $\phi$ with
the quintessence field, and contrary to Gardner also assumed 
$f>10^3 M_{Pl}$ in accordance with \citet{OP1}.
\citet{CNP} also identified $\phi$ with the quintessence and concluded 
that $f\sim 10^5 m_{Pl}$ to fit the QSO data. 
However, this result did not comply with the Oklo constraint, 
and Copeland et al. proposed that a photon momentum
dependence of $f$ around $10\,$MeV may suppress the effects of dynamical
$\alpha$ in nuclear reactions.

In the present paper we propose yet another analysis of the 
implications of the results of Murphy et al.
for a dynamical fine structure constant. In particular we assume
$m> 10^{-28}\,$eV for the mass of the scalar field $\phi$
generating the evolution of $\alpha$. Under this assumption 
a very weakly coupled field behaves like pressureless dust
ever since dust domination,
even though it may not satisfy the usual thermal dust 
condition\footnote{This general result that non-thermal
coherent oscillations behave like cold dark matter was
observed for the first time in axion physics \citep{AS,DF,PWW}.}
 $m\gg T$. The virtue of $m> 10^{-28}\,$eV for our
present analysis is that under this condition
we can use the late time behavior of $\phi(t)$ for $t\gg m^{-1}$
to characterize the evolution of $\phi$ ever since radiation--dust
equality.
 Furthermore, our ignorance about evolution of $\phi$ during radiation
domination can be collected in a single parameter\footnote{
$\xi\simeq 1$ would imply that $\phi$ is a dominant
cold dark matter component. This possibility was pointed out 
for heavy dilatons ($m\gg T$, dilaton {\sc wimp}s) 
by \citet{GV} and by \citet{DV}, and for oscillations
of light dilatons ($T\ge m\gg t^{-1}$) by \citet{rdmpla}.} 
$\xi=(\varrho_\phi/\varrho_m)^{1/2}$, and we perform a least squares
fit of the time evolution of $\alpha$ in our model to the
$\alpha_z$ values reported by \citet{murphy2}.

This explores a completely different mass range than \citet{gardner}.
 For $mt\gg 1$ the mass term generates temporal and spatial fluctuations
of $\phi$
at scales $m^{-1}$, but the Hubble expansion damps these oscillations
$\propto t^{-1}$, such that the amplitude of these oscillations
is well below current observational limits from laboratory based search 
experiments for variable $\alpha$.
 Furthermore, with $mt\gg 1$ we will be able to use a virial theorem
to eliminate $m$ from the long term variation of $\phi$. The fit
of the long term behavior of $\alpha(t)$ derived in Sec. \ref{sec:phit}
to the data of Murphy et al. then allows us to estimate
the parameter $f/\xi$.

 For cosmological parameters we rely on the recent evaluation
from the Wilkinson Microwave Anisotropy Probe (WMAP) 
and the Sloan Digital Sky Survey (SDSS) \citep{wmap,SDSS}.
We use in particular the values from the ``vanilla lite" model
\citep{SDSS}:
\begin{equation}\label{eq:vanilla}
\Omega_\Lambda=0.707^{+0.031}_{-0.039},
\quad
h=0.708^{+0.023}_{-0.024},
\end{equation}
\[
t_0=13.40^{+0.13}_{-0.12}\,\mathrm{Gyr}.
\]
We included the errors given by \citet{SDSS} for illustration,
but do not use them for error propagation. They are negligible
compared to the uncertainties in the $\alpha$ values for the
QSO absorption systems, which generate a $1\sigma$ uncertainty of about
22\%, see Eq. (\ref{eq:slope}) below. 

The dependence of the WMAP CMB results on $\alpha$ is relatively
weak in that it complies with $0.95<\alpha_{dec}/\alpha_0<1.02$
\citep{rocha}. The dynamical evolution of $\alpha$ calculated
below implies that at the time of decoupling
$0>(\alpha_{dec}-\alpha)/\alpha>-1\times 10^{-4}$, 
such that at this stage we can safely use WMAP results
on cosmological parameters for the determination of $f/\xi$.

Sec. \ref{sec:phit} recalls the relevant features of the dynamical 
evolution of a scalar field in an expanding universe and includes 
a virial theorem
that will be useful in the analysis of dynamical gauge couplings.

Our main result in Sec. \ref{sec:at} is an equation for the evolution
of $\alpha(t)$ from the $\phi$-$\gamma$ coupling,
and the fit to the results of Murphy et al. in Sec. \ref{sec:fitf}
yields $f/\xi$. In Sec. \ref{sec:oklo} we will compare the time
evolution of $\alpha$ in our model with the Oklo and meteorite constraints,
and Sec. \ref{sec:conc} contains our conclusions.

\section{The cosmological evolution of $\phi$}\label{sec:phit}

With $mt\gg 1$ the mass parameter induces spatial and temporal
fluctuations of $\phi$, and therefore of $\alpha$.
One might hope to use this to determine $m$ from a Fourier
decomposition of observations of $\alpha$ over cosmological
distances. Our primary interest here is the coupling scale $f$
of the scalar field to photons, and the strategy is to use
a fit of the long term evolution of $\phi$ in the expanding
universe to the time variation of $\alpha$ reported by Murphy 
et al. At this stage this allows us to infer a value for $f/\xi$.

$\phi$ is usually assumed to have at most extremely weak
matter couplings, and the long term evolution of
very weakly coupled helicity states follows
\begin{equation}\label{eq:motphi}
\ddot{\phi}(t)+3H(t)\dot{\phi}(t)+m^2\phi(t)=0,
\end{equation}
with a corresponding evolution of the comoving energy density
$\sqrt{-g}\varrho_\phi=a^3(\dot{\phi}^2+m^2\phi^2)/2$
\begin{equation}\label{eq:dEdt}
\frac{1}{2}\partial_0\left(a^3\dot{\phi}^2
+a^3m^2\phi^2\right)
=\frac{3}{2}Ha^3\left(
m^2\phi^2-\dot{\phi}^2\right).
\end{equation}
Note that in Eq. (\ref{eq:motphi}) we used the convention $\phi_0=0$,
cf. (\ref{eq:V}).

The solution of Eq. (\ref{eq:motphi}) for $m=0$,
\[
\partial_0\phi(t)\propto a^{-3}(t)
\]
implies $\varrho_\phi(t)\propto a^{-6}(t)$, as appropriate
for the ultrahard fluid component generated by massless weakly 
coupled helicity states (see e.g. \citet{rdplb2}).

The solution of Eq. (\ref{eq:motphi}) for $m>0$ and $a\sim t^{2/3}$
\[
\phi(t)=\frac{1}{\sqrt{t}}\left(
AJ_{\frac{1}{2}}(mt)+BY_{\frac{1}{2}}(mt)
\right)
\]
has asymptotics for $mt\gg 1$:
\[
\phi(t)\propto
\frac{1}{t}
\cos(mt+\varphi),
\]
\begin{equation}\label{eq:phia}
\varrho_\phi\propto t^{-2}
\propto a^{-3}.
\end{equation}
This means that at late times $p_\phi\simeq 0$ since
\[
\frac{d}{dt}\left(\varrho_\phi a^3\right)\simeq 0,
\]
just as for thermalized non-relativistic matter, but 
here even for $m\le T$. 

Eq. (\ref{eq:motphi}) can be used to express
the difference of
the comoving kinetic and potential energy densities 
as a time derivative
\begin{equation}\label{eq:vt1}
2a^3\mathcal{H}_{kin}-2a^3\mathcal{H}_{pot}
=a^3\dot{\phi}^2-m^2a^3\phi^2
=\frac{d}{dt}(a^3\phi\dot{\phi}).
\end{equation}
This implies for the time limit
\[
\overline{\mathcal{H}}=\lim_{\tau\to\infty}\frac{1}{\tau}
\int_0^\tau\! dt\,\mathcal{H}(t)
\]
a virial theorem
\[
\overline{a^3\mathcal{H}_{kin}}
=\overline{a^3\mathcal{H}_{pot}},
\]
\begin{equation}\label{eq:vt2}
\overline{a^3\varrho_\phi}\equiv
\overline{a^3\mathcal{H}}
=m^2\overline{a^3\phi^2}=\overline{a^3\dot{\phi}^2}.
\end{equation}
However, note that at late times $a^3\phi\dot{\phi}\propto t^0$,
and therefore Eq. (\ref{eq:vt1}) also yields 
\begin{equation}\label{eq:kp}
\mathcal{H}_{kin}\simeq\mathcal{H}_{pot},
\end{equation}
i.e.
\begin{equation}\label{eq:vtend1}
|\dot{\phi}|\simeq m|\phi|,
\end{equation}
\begin{equation}\label{eq:vtend2}
\varrho_\phi\simeq \dot{\phi}^2.
\end{equation}
This relation can be used to trade the mass dependent $|\dot{\phi}|$ 
for the energy density $\varrho_\phi$ in the late time evolution
$(t\gg m^{-1})$ of $\alpha$.

\section{The cosmological evolution of $\alpha$}\label{sec:at}

 From Eq. (\ref{eq:Qlin}) we have 
with $\alpha\equiv\alpha_0\equiv\alpha(t_0)$:
\[
\frac{\alpha(t)}{\alpha}=\frac{e^2(\phi(t))}{e^2}
\simeq 1-\frac{\phi}{f},
\]
and with (\ref{eq:vtend2})
\begin{equation}\label{eq:da/a1}
\frac{|\dot{\alpha}(t)|}{\alpha}\simeq\frac{|\dot{\phi|}}{f}
\simeq\frac{\sqrt{\varrho_\phi}}{f}.
\end{equation}
We know from Eq. (\ref{eq:phia}) and Eq. (\ref{eq:rhomat})
in the Appendix that
\[
\varrho_\phi(t)=\varrho_{\phi,0}\left(\frac{a_0}{a(t)}\right)^3
=\frac{\varrho_{\phi,0}}{\varrho_{m,0}}\varrho_m(t)
=\frac{\varrho_{\phi,0}}{\varrho_{m,0}}
\frac{\Lambda}{\sinh^2(t/\tau)}.
\]
We denote the contribution from the dynamical gauge coupling 
to the matter density by
$\xi^2=\varrho_{\phi,0}/\varrho_{m,0}$, and find for the rate
of change of the fine structure constant
\begin{equation}\label{eq:da/a2}
\frac{|\dot{\alpha}(t)|}{\alpha}
\simeq
\frac{\xi}{f}
\frac{\sqrt{\Lambda}}{\sinh(t/\tau)}.
\end{equation}
Integration yields
\begin{equation}\label{eq:alphat}
\ln\left(\frac{\alpha(t)}{\alpha(t_0)}\right)=
\xi\frac{2m_{Pl}}{\sqrt{3}f}
\ln\left(
\frac{\tanh(t/2\tau)}{\tanh(t_0/2\tau)}\right).
\end{equation}
The time constant is with $\Lambda=0.707\varrho_c$
\[
\tau=\frac{2m_{Pl}}{\sqrt{3\Lambda}}
=5.250\times 10^{32}\,\mbox{eV}^{-1}
=3.455\times 10^{17}\,\mbox{s}=10.95\,\mbox{Gyr}.
\]

To match Eq. (\ref{eq:alphat}) 
to the data from \citet{murphy2} we also need the 
redshift-time relation for $z\ll 10^3$
from (\ref{eq:scale2})
\begin{eqnarray}\label{eq:tz}
t_z&=&\tau\sinh^{-1}\!\left(
\frac{\sinh(t_0/\tau)}{
(1+z)^{1.5}}
\right)\\
\nonumber
&=&10.95
\times
\sinh^{-1}\!\left(
\frac{1.553}{(1+z)^{1.5}}
\right)
\,\mbox{Gyr}.
\end{eqnarray}

Eqs. (\ref{eq:alphat}) and (\ref{eq:tz}) allow for
a determination of the parameter $f/\xi$.

\section{The coupling scale}\label{sec:fitf}

The parameter $f/\xi$ was determined from a fit of the Eqs.
(\ref{eq:alphat},\ref{eq:tz}) to the $\alpha(z)$ values reported
by \citet{murphy2}. We set
\[
y_i=\ln\left(\frac{\alpha(t_i)}{\alpha(t_0)}\right),
\]
\[
x_i=
\ln\left(
\frac{\tanh(t_i/2\tau)}{\tanh(t_0/2\tau)}\right),
\]
and the minimal variance
\[ 
\chi^2=\sum_i\left(\frac{y_i-sx_i}{\delta y_i}\right)^2
\]
in the fit of (\ref{eq:alphat}) occurs for a slope
\[
s=\xi\frac{2m_{Pl}}{\sqrt{3}f}=
\frac{\sum_i x_iy_i/\delta y_i^2}{
\sum_j x_j^2/\delta y_j^2}
\]
with variance
\[
\sigma_s^2=\sum_i\delta y_i^2\left(\partial s/\partial y_i\right)^2
=\left(\sum_i x_i^2/\delta y_i^2\right)^{-1}.
\]
The errors in $\ln(\alpha_z/\alpha_0)$ are related
to the errors in $\Delta\alpha/\alpha_0$ through
\[
\delta y_i=\frac{\delta\Delta\alpha}{\alpha_0+\Delta\alpha}.
\]

This method yields a slope 
\begin{equation}\label{eq:slope}
s=(5.443\pm 1.174)\times 10^{-6}
\end{equation}
corresponding to a coupling parameter
\begin{equation}\label{eq:fvalue}
\frac{f}{\xi}=2.12^{+0.58}_{-0.37}\times 10^5m_{Pl}.
\end{equation}
The variance per degree of freedom is $\chi^2_{dof}=1.067$.

The resulting asymptotic equilibrium value of the fine structure
constant is
\[
\lim_{t\to\infty}\alpha(t)/\alpha(t_0)=\tanh(t_0/2\tau)^{-s}
\simeq 1+(3.3\pm 0.7)\times 10^{-6}.
\]
The current rate of change of $\alpha$ from Eqs. 
(\ref{eq:alphat},\ref{eq:slope})
\begin{equation}\label{eq:rate}
\frac{\dot{\alpha}}{\alpha}=\frac{s}{4\tau\sinh(t_0/\tau)}
=(4.0\pm 0.9)\times 10^{-17}\,\mathrm{yr}^{-1}
\end{equation}
is within the bounds from current atomic clock experiments
\citep{marion}:
\[
-2.0\times 10^{-16}\,\mathrm{yr}^{-1}\le\frac{\dot{\alpha}}{\alpha}
\le 1.2\times 10^{-16}\,\mathrm{yr}^{-1}.
\]

\section{Comparison with the Oklo and meteorite constraints}\label{sec:oklo}

With the cosmological parameters (\ref{eq:vanilla}) and Eq. (\ref{eq:tz})
the closest QSO absorption system used in \citet{murphy2} corresponds to a 
distance of about 2.7 billion light years. A well known more recent 
constraint on variations of $\alpha$ over cosmological time scales comes 
from isotope abundances in the natural Oklo reactor, which had
been active about 1.8 billion years ago \citep{oklo1,DD,fujii}.

The recent evaluation by \citet{fujii} yields a bound 
\begin{equation}\label{eq:oklo}
\frac{\alpha_{Oklo}-\alpha}{\alpha}=-(0.8\pm 1.0)\times 10^{-8},
\end{equation}
whereas insertion of (\ref{eq:fvalue}) into (\ref{eq:alphat}) yields
\[
\frac{\alpha_{Oklo}-\alpha}{\alpha}=-(6.38^{+1.37}_{-1.38})\times 10^{-7}.
\]
The situation appears to be different with the ${}^{187}$Re constraints,
which limit the evolution of $\alpha$ over the last 4.6\,Gyr 
\citep{meteor1,freeman,OP2}: 
The most recent constraint from \citet{OP2} is
\[
\frac{\alpha_{4.6}-\alpha}{\alpha}=-(0.8\pm 0.8)\times 10^{-6},
\]
while Eqs. (\ref{eq:fvalue},\ref{eq:alphat}) yield
\[
\frac{\alpha_{4.6}-\alpha}{\alpha}=-(1.95\pm 0.42)\times 10^{-6}.
\]
\citet{OP1} and \citet{gardner} could fit their externally
driven models for the evolution of $\alpha$ to both the QSO
data and the Oklo constraint, whereas \citet{luis}, \citet{CNP}
and we find a lower value of $\alpha$. However, \citet{mota} have 
recently pointed out that the local variation of $\alpha$ in virialized 
overdensities like our own can relax the Oklo constraint by a factor
10-100, because on the one hand a dynamical $\alpha$ would be 
expected to have a higher value in overdensities, while on the other
hand virialization of overdensities slows down the evolution of $\alpha$.
The qualitative picture emerging from this is that in overdensities
$\alpha$ evolves from a higher initial value after virialization, but at
slower pace, whence it approaches again the value in  the low-density
background universe. This can explain discrepancies 
between astrophysical and geochemical observations. 

\section{Conclusions}\label{sec:conc}

We have found the equation (\ref{eq:alphat}) for the dynamical time
evolution of $\alpha$ due to the coupling $-(\phi/4f)F^{\mu\nu}F_{\mu\nu}$
of electromagnetic fields to a very weakly coupled massive scalar field
with $mt\gg 1$. A fit of this equation to the quasar absorption data
reported by \citet{murphy2} yields the value (\ref{eq:fvalue}), where
$\xi=\sqrt{\varrho_{\phi,0}/\varrho_{m,0}}$ parametrizes the contribution
of the scalar field $\phi$ to the matter density.

Within this model the evolution of $\alpha$ reported by Murphy et al.
appears to be slow due to a small coefficient 
$s=(5.443\pm 1.174)\times 10^{-6}$
in Eq. (\ref{eq:alphat}): The fine-structure constant varied so little
since $z=3.66$ because the $\phi$ abundance is small and the
$\phi$-$\gamma$ coupling is very weak and presumably of subgravitational 
strength, in agreement with the analyses of \citet{OP1,luis}
and \citet{CNP}.

Our coherent oscillation model for dynamical $\alpha$ (and other
self-driven models of dynamical $\alpha$ \citep{luis,CNP})
still seems to predict a too small value of $\alpha$ at the time 
when the Oklo natural reactor was active.
However, as \citet{mota} have pointed out, at low redshift
predictions of faster evolution of $\alpha$ from
astrophysical observations are to be expected due to spatial variations
in the presence of local overdensities. Note that this does not affect 
the numerical results (\ref{eq:slope},\ref{eq:fvalue},\ref{eq:rate}) 
(or the corresponding results of \citet{luis} and \citet{CNP}), since 
the resulting discrepancy of $\alpha_0$ on Earth
and far away from virialized objects is smaller than the number of 
significant figures reported\footnote{We should emphasize that the 
spatial variation effect discussed by \citet{mota} cannot be used as an 
argument against the models of \citet{OP1} and \citet{gardner}, since 
externally driven models also rather tend to predict a smaller $\alpha$ 
than the value inferred from the Oklo data. Externally driven models 
only seem to provide more leeway to fit time evolutions of $\alpha$.}.

\section*{Acknowledgments}

This work was supported through the Natural Sciences and Engineering
Research Council of Canada.

\appendix

\section[]{The scale factor in the $\Lambda$CDM universe}\label{sec:app1}

The evolution of the scale factor in a spatially flat 
$\Lambda$CDM universe
follows from direct integration of the corresponding Friedmann equation
\begin{equation}\label{eq:fld}
\frac{\dot{a}^2}{a^2}=\frac{\varrho_{m}+\Lambda}{3m_{Pl}^2}
\end{equation}
after insertion of
\[
\varrho_{m}(t)=\varrho_{m,0}\left(
\frac{a_0}{a(t)}\right)^3.
\]
Integration from $t_0$ to $t$ yields
\[
\frac{\sqrt{3\Lambda}}{2m_{Pl}}(t-t_0)
=\ln\!\left(
\frac{\sqrt{\Lambda a^3}+\sqrt{\Lambda a^3+\varrho_{m,0}a_0^3}}
{\sqrt{\Lambda a_0^3}+\sqrt{(\Lambda+\varrho_{m,0})a_0^3}}
\right).
\]
Solving for the scale factor yields 
\begin{eqnarray}\nonumber
\left(\frac{a}{a_0}\right)^{3/2}
&=&\exp\!\left(\frac{t-t_0}{\tau}\right)\\
&&+\frac{\varrho_{m,0}}{\Lambda+\sqrt{\Lambda^2+\Lambda\varrho_{m,0}}}
\sinh\!\left(\frac{t-t_0}{\tau}\right),
\label{eq:scale1}
\end{eqnarray}
with the time constant
\begin{equation}\label{eq:taudef}
\tau=\frac{2m_{Pl}}{\sqrt{3\Lambda}}.
\end{equation}

We can simplify our result (\ref{eq:scale1}) because the highest redshift
$z=3.66$ used in the analysis still corresponds to an age
$t(z)\simeq 1.7\,\mathrm{Gyr}\gg t_{eq}$ 
much larger than the time 
$t_{eq}\simeq 1.3\times 10^5\,$yr
of matter radiation equality. At times $\gg t_{eq}$ the modification
of the time evolution of the scale factor during the very early
radiation dominated era can be neglected, and one can integrate
Eq. (\ref{eq:fld}) from $t_1=0$ and still get an extremely good
approximation for $t\gg t_{eq}$. This yields
\begin{equation}\label{eq:scale2}
\frac{a}{a_0}
=\left(\frac{\sinh(t/\tau)}{\sinh(t_0/\tau)}
\right)^{2/3}
\end{equation}
and
\begin{equation}\label{eq:rhomat}
\varrho_m=\frac{\Lambda}{\sinh^2(t/\tau)}.
\end{equation}

\bsp

\label{lastpage}

\end{document}